\documentclass[amsmath,aps,pra,reprint,longbibliography,superscriptaddress]{revtex4-2}
\usepackage{graphicx, hyperref}
\usepackage[caption=false]{subfig}
\allowdisplaybreaks[1] 

\begin{document}


\newcommand{\ket}[1]{\vert #1 \rangle}
\newcommand{\bra}[1]{\langle #1 \vert}
\newcommand{\braket}[2]{\langle #1 \vert #2 \rangle}
\newcommand{\ketbra}[2]{\lvert #1 \rangle \langle #2 \rvert}
\newcommand{\absv}[1]{\lvert #1 \rvert}
\newcommand{\fref}[1]{Fig.~\ref{#1}}
\newcommand{\Fref}[1]{Figure~\ref{#1}}
\newcommand{\sref}[1]{Sec.~\ref{#1}}
\newcommand{\tref}[1]{Table.~\ref{#1}}


\title{Constant-Time Quantum Search with a Many-Body Quantum System}

\author{Benjamin DalFavero}
    \altaffiliation{Present address: Department of Computational Mathematics, Science, and Engineering, Michigan State University, East Lansing, MI 48824, USA}
    \affiliation{Department of Physics, Creighton University, Omaha, NE 68178, USA}
\author{Alexander Meill}
    \affiliation{Department of Physics, University of California, San Diego, La Jolla, CA 92093, USA}
\author{David A. Meyer}
    \affiliation{Department of Mathematics, University of California, San Diego, La Jolla 92093, CA, USA}
\author{Thomas G. Wong}
    \affiliation{Department of Physics, Creighton University, Omaha, NE 68178, USA}
\author{Jonathan P. Wrubel}
    \affiliation{Department of Physics, Creighton University, Omaha, NE 68178, USA}

\begin{abstract}
    The optimal runtime of a quantum computer searching a database is typically cited as the square root of the number of items in the database, which is famously achieved by Grover's algorithm. With parallel oracles, however, it is possible to search faster than this. We consider a many-body quantum system that naturally effects parallel queries, and we show that its parameters can be tuned to search a database in constant time, assuming a sufficient number of interacting particles. In particular, we consider Bose-Einstein condensates with pairwise and three-body interactions in the mean-field limit, which effectively evolve by a nonlinear Schr\"odinger equation with cubic and quintic nonlinearities. We solve the unstructured search problem formulated as a continuous-time quantum walk searching the complete graph in constant time. Depending on the number of marked vertices, however, the success probability can peak sharply, necessitating high precision time measurement to observe the system at this peak. Overcoming this, we prove that the relative coefficients of the cubic and quintic terms can be tuned to eliminate the need for high time-measurement precision by widening the peak in success probability or having it plateau. Finally, we derive a lower bound on the number of atoms needed for the many-body system to evolve by the effective nonlinearity.
\end{abstract}

\maketitle


\section{\label{sec:intro}Introduction}

Grover's quantum search algorithm famously searches an unstructured database of size $N$ in at most $O(\sqrt{N})$ timesteps \cite{Grover1996}. Bennett, Bernstein, Brassard, and Vazirani \cite{Bennett1997} showed that a quantum computer needs at least $\Omega(\sqrt{N})$ steps to perform the search, and thus Grover's algorithm is optimal.

All this, however, assumes access to a single oracle. Zalka \cite{Zalka1999} considered parallel oracles and showed that the number of oracles $S$ times the square of the runtime $T$ is lower bounded by the number of items in the database, i.e.,
\begin{equation}
    \label{eq:Zalka}
    ST^2 = \Omega(N).
\end{equation}
When there is one oracle, $S = 1$, and then taking the square root of both sides of \eqref{eq:Zalka}, we get $T = \Omega(\sqrt{N})$, in agreement with Bennett et al. As the number of oracles $S$ increases, \eqref{eq:Zalka} implies that the lower bound on the runtime $T$ decreases, opening the possibility for a quantum computer to search more quickly than $O(\sqrt{N})$.

\begin{figure}
	\includegraphics{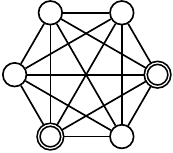}
	\caption{\label{fig:complete} The complete graph of $N = 6$ vertices. Of them, $k = 2$ are marked (indicated by double circles) and $N-k=4$ are unmarked (indicated by single circles).}
\end{figure}

Formulated as a continuous-time quantum walk \cite{FG1998-analog,CG2004}, the unstructured database corresponds to the complete graph of $N$ vertices, and the goal is to find one of $k$ vertices ``marked'' by an oracle. An example is shown in \fref{fig:complete}. In this formulation, the $N$ vertices of the graph correspond to the computational basis states $\ket{1}, \ket{2}, \dots, \ket{N}$, and the system is a quantum mechanical walker that evolves in a superposition of these basis states. The walker begins in a uniform superposition over all $N$ vertices, and it evolves by Schr\"odinger's equation,
\[ i \frac{\partial \psi}{\partial t} = H \psi, \]
with some appropriate Hamiltonian $H$ that includes a term that effects the quantum walk and a term that acts as the oracle. Note we have also set $\hbar = 1$ without affecting our asymptotic results. At a time $O(\sqrt{N/k})$ \cite{Wong10}, the walker is in a uniform superposition over the marked vertices, and so the probability of finding the walker at a marked vertex is 1. When there is a single marked vertex, $k = 1$, and the runtime is $O(\sqrt{N})$, as expected.

For a many-body quantum system undergoing a multiparticle quantum walk \cite{Childs2013}, each particle interacts with the oracle in parallel. For example, in a Bose-Einstein condensate \cite{Bose1924,Einstein1924,Einstein1925} in an optical lattice undergoing a quantum walk between lattice sites, each boson interacts with the oracle at the marked sites. This is equivalent to having many oracles, and so Zalka's lower bound on the runtime \eqref{eq:Zalka} holds. This raises the possibility that a many-body quantum system can search faster than $O(\sqrt{N})$, at the expense of needing many particles.

Although quantum mechanics is linear, several many-body quantum systems evolve with effective nonlinearities. A famous example is a Bose-Einstein condensate, which for low temperatures, high numbers of particles, and pairwise contact interactions is asymptotically described by the Gross-Pitaevskii equation \cite{Gross1961,Pitaevskii1961}, a Schr\"odinger-type equation with a cubic nonlinearity $\absv{\psi}^2\psi$:
\begin{equation}
    \label{eq:cubic}
    i \frac{\partial \psi}{\partial t} = ( H - g \absv{\psi}^2 ) \psi,
\end{equation}
where $g$ is a real parameter dictating the strength of the nonlinearity. Since $\absv{\psi}^2$ is the probability density, the nonlinearity acts more strongly in regions with higher probability density. Thus, we can think of $g \absv{\psi}^2$ as a nonlinear ``self-potential.''

Under certain conditions, the effectively nonlinear evolution of a Bose-Einstein condensate contains an additional quintic nonlinearity $|\psi|^4\psi$, such as for a Bose-Einstein condensate with both pairwise and three-body interactions \cite{Gammal2000} or for a Bose-Einstein condensate confined in a pseudo-1D potential \cite{Trallero2013}. In such systems with both cubic and quintic nonlinearities, the effective Schr\"odinger-type equation is
\begin{equation}
    \label{eq:cubic-quintic}
    i \frac{\partial \psi}{\partial t} = \left[ H - g \left( \absv{\psi}^2 - h \absv{\psi}^4 \right) \right] \psi,
\end{equation}
where $h$ is a non-negative parameter dictating the relative strength of the quintic term.

Algorithmic consequences of computing with various nonlinearities have been explored in a variety of contexts, such as for solving NP-complete and \#P problems \cite{AL1998}, for state discrimination \cite{Childs2016}, and for searching \cite{Wong3,Wong4,Childs2016,DiMolfetta2020}. In this paper, we focus on spatial search on the complete graph \cite{CG2004} using a nonlinear quantum walk, which was explored in \cite{Wong3} using the cubic nonlinearity \eqref{eq:cubic}, in \cite{Wong4} with the cubic-quintic nonlinearity \eqref{eq:cubic-quintic} for the case where $h = 1$, and in \cite{Wong4} using a loglinear nonlinearity, in each case with a time-dependent hopping rate. In this paper, we will consider the cubic-quintic nonlinearity \eqref{eq:cubic-quintic} with general $h$ and show that previously unseen behavior arises, including evolution that asymptotically approaches a steady state rather than evolving periodically (which is possible since the evolution is non-autonomous), and the ability to tune the algorithm to reduce the time-measurement precision needed for the algorithm, which enables searching in constant time without increased time-measurement precision.

\begin{figure}
\begin{center}
	\subfloat[]{
		\includegraphics{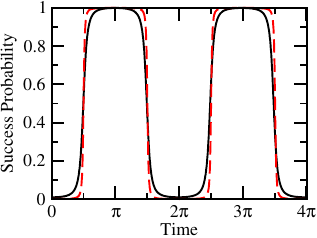}
		\label{fig:prob-h1-k1}
	}
 
	\subfloat[]{
		\includegraphics{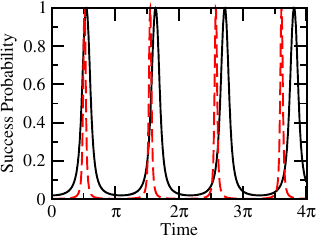}
		\label{fig:prob-h1-k2}
	}
	\caption{\label{fig:prob-h1}Success probability as a function of time for search on the complete graph of $N$ vertices using the cubic-quintic nonlinearity \eqref{eq:cubic-quintic} with $h = 1$ and (a) $k = 1$ and (b) $k = 2$ marked vertices. The solid black curve is $N = 100$, and the dashed red curve is $N = 1000$. The nonlinear coefficient is $g = N-1$.}
\end{center}
\end{figure}

To illustrate the behaviors that have been previously seen, and the role that time-measurement precision plays, let us consider an example of searching with the cubic-quintic nonlinearity with $h = 1$, as explored in \cite{Wong4}.  In \fref{fig:prob-h1}, we have plotted the probability of measuring the walker at a marked vertex (henceforth called the ``success probability'') against time $t$. In \fref{fig:prob-h1-k1}, the number of marked vertices is $k = 1$. The solid black curve shows the success probability when searching the complete graph with $N = 100$ vertices, and it starts at $1/100$ and reaches a value of $1$ at time $t = \pi$ before returning to its initial value. The dashed red curve shows the success probability when searching the complete graph with $N = 1000$ vertices, and it starts at $1/1000$ and also reaches a value of $1$ at time $t = \pi$. Thus, regardless of $N$, if we measures the position of the walker at time $t = \pi$, we are guaranteed to find it at the marked vertex, which accomplishes the search. Since the runtime is independent of $N$, this is a constant-runtime algorithm.

Next, let us consider when there are $k = 2$ marked vertices. Intuitively, one might imagine that this search problem is less challenging because with two marked vertices, it should be easier to find one of them. With the cubic-quintic nonlinear quantum walk with $h = 1$, however, it can be more difficult. In \fref{fig:prob-h1-k2}, we plot search with $k = 2$ marked vertices, again with the solid black and dashed red curves corresponding to complete graphs with $N = 100$ and $N = 1000$ vertices, respectively. Now, the success probability peaks around $t = \pi/2$, but the peak is very narrow, and it gets more narrow for larger graphs, meaning a more precise clock is needed in order to measure the position of the walker at just the right time. Using an insufficiently precise clock could lead to accidentally measuring the walker as being at an unmarked vertex, meaning that the algorithm must be repeated. For larger values of $k$, the success probability also peaks sharply.

\begin{figure*}
\begin{center}
	\subfloat[$h = 1$]{
		\includegraphics{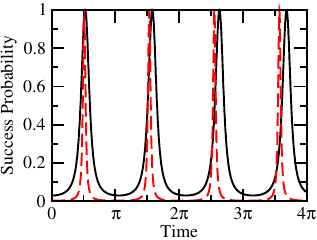}
		\label{fig:prob-k3-h1}
	}
	\subfloat[$h = 2$]{
		\includegraphics{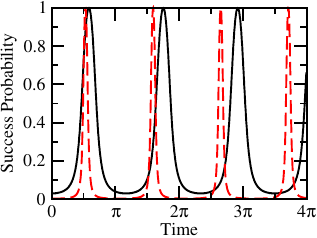}
		\label{fig:prob-k3-h2}
	} 
	\subfloat[$h = 2.9$]{
		\includegraphics{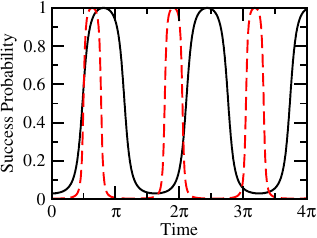}
		\label{fig:prob-k3-h2.9}
	}
 
	\subfloat[$h = 2.99$]{
		\includegraphics{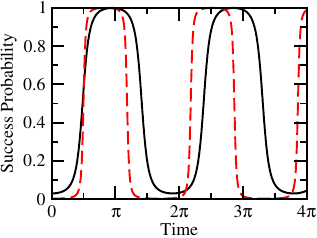}
		\label{fig:prob-k3-h2.99}
	}
	\subfloat[$h = 3$]{
		\includegraphics{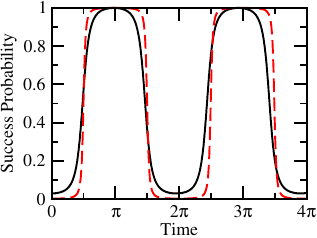}
		\label{fig:prob-k3-h3}
	}
	\subfloat[$h = 3.008$]{
		\includegraphics{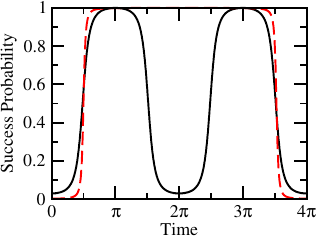}
		\label{fig:prob-k3-h3.008}
	}
 
	\subfloat[$h = 3.0091$]{
		\includegraphics{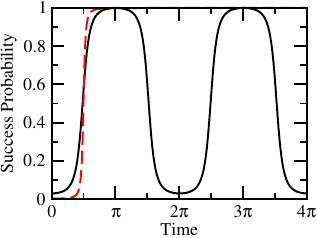}
		\label{fig:prob-k3-h3.0091}
	}
	\subfloat[$h = 3.08$]{
		\includegraphics{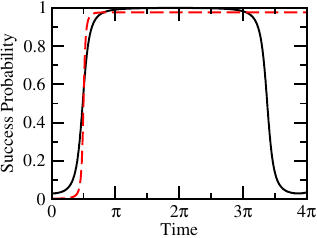}
		\label{fig:prob-k3-h3.08}
	}
	\subfloat[$h = 3.091$]{
		\includegraphics{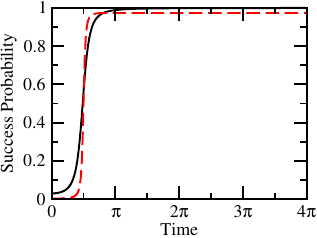}
		\label{fig:prob-k3-h3.091}
	}
 
	\subfloat[$h = 3.3$]{
		\includegraphics{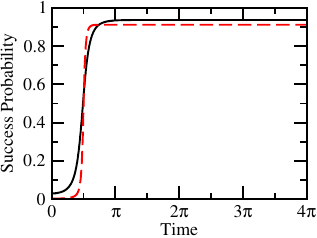}
		\label{fig:prob-k3-h3.3}
	}
	\subfloat[$h = 4$]{
		\includegraphics{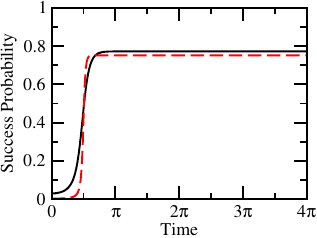}
		\label{fig:prob-k3-h4}
	}
	\subfloat[$h = 5$]{
		\includegraphics{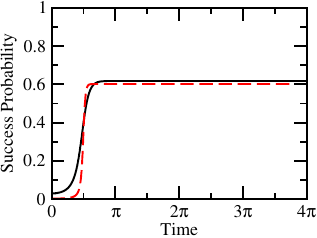}
		\label{fig:prob-k3-h5}
	}
	\caption{\label{fig:prob-k3} Success probability as a function of time for search on the complete graph of $N$ vertices and $k = 3$ marked vertices using the cubic-quintic nonlinearity \eqref{eq:cubic-quintic} with $g = N-1$ and various values of $h$. The solid black curve is $N = 100$, and the dashed red curve is $N = 1000$.}
\end{center}
\end{figure*}

In this paper, we show how to eliminate the increased time-measurement precision by tuning $h$ in \eqref{eq:cubic-quintic}. Besides its algorithmic benefits, it is also motivated by physical systems, where the cubic and quintic terms may have different strengths \cite{Gammal2000,Trallero2013}. As an example of the algorithmic advantage, the success probability when searching by \eqref{eq:cubic-quintic} with $k = 3$ is plotted in \fref{fig:prob-k3} with different values of $h$ in each subfigure. In all of these, the solid black curves correspond to searching the complete graph of $N = 100$ vertices, and the red dashed curves correspond to $N = 1000$. We see that when $h < k$ in \fref{fig:prob-k3-h1} and \fref{fig:prob-k3-h2}, the success probability peaks sharply at time $\pi/2$, similar to \fref{fig:prob-h1-k2}. As $h$ nears $k$ in \fref{fig:prob-k3-h2.9} and \fref{fig:prob-k3-h2.99}, the peaks begin to stretch out, and when $h = k$ in \fref{fig:prob-k3-h3}, the success probably has a wide peak at time $\pi$, similar to \fref{fig:prob-h1-k1}. When $h$ is slightly larger than $k$ in \fref{fig:prob-k3-h3.008}, the red dashed curve corresponding to $N = 1000$ has stretched out its peak considerably, while the solid black curve corresponding to $N = 100$ has only stretched out marginally. With a slightly larger value of $h$ in \fref{fig:prob-k3-h3.0091}, the red dashed curve now plateaus at a success probability of 1. In \fref{fig:prob-k3-h3.08} and \fref{fig:prob-k3-h3.091}, the solid black curve widens and then plateaus as well. For subsequently larger values of $h$ in \fref{fig:prob-k3-h3.3} through \fref{fig:prob-k3-h5}, the plateaus occur at lower and lower success probabilities. Note that the plateauing success probability in \fref{fig:prob-k3-h3.0091} through \fref{fig:prob-k3-h5} is a new behavior not previously seen in \cite{Wong3,Wong4}; in those studies, all the nonlinearities resulted in periodic success probabilities that reached 1. The flat success probability means the measurement time does not need to be very precise. Furthermore, although the success probability can plateau at a value less than one, it is asymptotically independent of $N$, and so on average, the algorithm only needs to be repeated a constant number of times before a marked vertex is found, and thus the overall algorithm runs in constant time.

In \sref{sec:spatial-search}, we review the nonlinear quantum walk algorithm for search on the complete graph. Then, in \sref{sec:peaks}, we derive analytically the criteria for regions where the success probability has a sharp peak, has a wide peak, or plateaus, as we numerically saw in \fref{fig:prob-k3}. Following, in \sref{sec:runtime}, we determine analytically what time the sharp peaks, wide peaks, and plateaus are reached, including proving the constant-time results suggested by \fref{fig:prob-k3}. In \sref{sec:resources}, we explore additional resources implied by the optimality of Grover's algorithm \cite{Zalka1999} for the nonlinear algorithm considered here, yielding a lower bound on the number of atoms needed for the many-body quantum system to evolve by the cubic-quintic nonlinearity. We conclude in \sref{sec:conclusion}.


\section{\label{sec:spatial-search}Nonlinear Spatial Search}

We begin our analysis by reviewing the nonlinear search algorithm of \cite{Wong4}. First, note that both \eqref{eq:cubic} and \eqref{eq:cubic-quintic} have the form of a nonlinear Schr{\"o}dinger equation 
\begin{equation}
	\label{eq:NLSE}
	i \frac{\partial \psi}{\partial t} = \left[ H - g f(\absv{\psi}^2) \right] \psi,
\end{equation}
where $f$ is some real-valued function. The cubic equation \eqref{eq:cubic} corresponds to $f(p) = p$, the previously analyzed cubic-quintic equation \eqref{eq:cubic-quintic} with $h = 1$ corresponds to $f(p) = p - p^2$, and the general cubic-quintic equation \eqref{eq:cubic-quintic} that is the focus of this paper corresponds to $f(p) = p - hp^2$. The nonlinear self-potential is $g f(|\psi|^2)$.

The system $\ket{\psi(t)}$ begins in a uniform superposition $\ket{s}$ over all $N$ vertices:
\begin{equation}
    \label{eq:s}
    \ket{\psi(0)} = \ket{s} = \frac{1}{\sqrt{N}} \sum_{i=1}^N \ket{i}.
\end{equation}
This state will evolve by \eqref{eq:NLSE} with
\begin{equation}
    \label{eq:search-hamiltonian}
    H = -\gamma N \ketbra{s}{s} - \sum_{i \in M} \ketbra{i}{i},
\end{equation}
where $\gamma$ is the jumping rate of the walker, and $M$ is the set of vertices marked by the oracle. $H$ is identical to the Hamiltonian for the linear quantum walk algorithm that searches the complete graph, introduced by Childs and Goldstone \cite{CG2004} in their Eq.~(10). The first term of \eqref{eq:search-hamiltonian} is equivalent to the discrete Laplacian for the complete graph, which is proportional to the kinetic energy of the walker that allows it to walk from vertex to vertex. The second term is the oracle query, which uniformly lowers the potential energy of marked states relative to unmarked states. In \eqref{eq:NLSE}, the nonlinear self-potential, adapted to on-site interactions in discretized space, becomes
\begin{equation}
    \label{eq:nonlinear-potential}
    g f(|\psi|^2) \to g \sum_{i=1}^{N} f {\left( \absv{\braket{i}{\psi (t)}}^2 \right)} \ketbra{i}{i}.
\end{equation}
Thus, the system evolves from \eqref{eq:s} by \eqref{eq:NLSE} with \eqref{eq:search-hamiltonian} and \eqref{eq:nonlinear-potential}.

Due to the symmetries present in the complete graph, all the marked vertices evolve with identical amplitudes, and all the unmarked vertices evolve with identical amplitudes. Thus, the state of the system can be expressed in terms of marked and unmarked subspaces:
\[ \ket{\psi (t)} = \alpha(t) \frac{1}{\sqrt{k}} \sum_{i \in M} \ket{i}    + \beta(t) \frac{1}{\sqrt{N-k}} \sum_{i \notin M} \ket{i}, \]
where $\alpha(0) = \sqrt{k/N}$ and $\beta(0) = \sqrt{(N-k)/N}$ correspond to the initial uniform superposition state. In this subspace, the success probability is $\absv{\alpha}^2$, and the ``failure probability'' (i.e., probability that the walker is measured to be at an unmarked vertex) is $\absv{\beta}^2 = 1 - \absv{\alpha}^2$. Since there are $k$ marked vertices that evolve with equal amplitudes, the probability of measuring the particle to be at any one particular marked vertex is $|\alpha|^2/k$. Similarly, since there are $N-k$ unmarked vertices that evolve with equal amplitudes, the probability of measuring the particle to be at any one particular unmarked vertex is $|\beta|^2/(N-k)$. Then, from \eqref{eq:nonlinear-potential}, apart from a factor of $g$, the strength of the nonlinear self-potential at each marked and unmarked vertex is given by $f_\alpha$ and $f_\beta$, respectively, where
\begin{equation}
    \label{eq:f-alpha-beta}
    f_\alpha = f {\left( \frac{\absv{\alpha}^2}{k} \right)}, \quad \text{and} \quad f_\beta = f {\left( \frac{\absv{\beta}^2}{N-k} \right)}.
\end{equation}
Using these definitions, the nonlinear Schr\"odinger equation \eqref{eq:NLSE} for the search algorithm becomes \cite{Wong4}
\begin{equation}
    \label{eq:eom}
    \frac{d}{dt}
    \begin{bmatrix}
        \alpha \\
        \beta
    \end{bmatrix}
    = i
    \begin{bmatrix}
        \gamma k + 1 + g f_\alpha & \gamma \sqrt{k} \sqrt{N-k} \\
        \gamma \sqrt{k} \sqrt{N-k} & \gamma (N-k) + g f_\beta
    \end{bmatrix}
    \begin{bmatrix}
        \alpha \\
        \beta
    \end{bmatrix}.
\end{equation}
As shown in \cite{Wong3,Wong4}, the jumping rate $\gamma$ should take a critical value of
\begin{equation}
	\label{eq:gamma-critical}
	\gamma_c = \frac{1}{N} \left[ 1 + g \left( f_\alpha - f_\beta \right) \right].
\end{equation}
Using this value of $\gamma$ and denoting the success probability $x(t) = \absv{\alpha(t)}^2$, it was proved in \cite{Wong4} that the evolution time $t$ at which the success probability reaches a value of $x$ is given by
\begin{equation}
	\label{eq:time}
	t = \frac{N}{2\sqrt{k}} \int_{k/N}^{x} \frac{1}{1 \!+\! g(f_\alpha \!-\! f_\beta)} \frac{1}{\sqrt{(1\!-\!x)(Nx\!-\!k)}} dx.
\end{equation}
As noted in the introduction, \cite{Wong4} only considered nonlinearities that resulted in the success probability reaching 1, so the runtime $t_*$ of the algorithm was defined as the time to reach a success probability of $1$, which is obtained by integrating \eqref{eq:time} from $x = k/N$ to $1$:
\begin{equation}
	\label{eq:runtime}
	t_* = \frac{N}{2\sqrt{k}} \int_{k/N}^{1} \frac{1}{1 \!+\! g(f_\alpha \!-\! f_\beta)} \frac{1}{\sqrt{(1\!-\!x)(Nx\!-\!k)}} dx.
\end{equation}
This is the time at which the walker will definitely be found at a marked vertex. For the $h = 1$ case, it was shown in \cite{Wong4} that when $g = N-1$, the success probability peaks in constant time, as in \fref{fig:prob-h1}. In \sref{sec:runtime}, we will prove that for general $h$, $g = N-1$ also causes the success probability to reach its plateau or peak in constant time, as shown in \fref{fig:prob-k3}.

From \cite{Wong3,Wong4}, the time-measurement precision is the width of the peak in the success probability around the points where $x = 1$, which can be found by finding the width of the peak at height $1-\epsilon$, which for a nonlinear Schr\"odinger equation of the form \eqref{eq:NLSE} is \cite{Wong4}
\begin{equation}
	\label{eq:width}
	\Delta t^{(0)} = \frac{2N}{1+g(\left. f_\alpha \right|_{x=1} - \left. f_\beta \right|_{x=1})} \sqrt{\frac{\epsilon}{k(N-k)}},
\end{equation}
where $\left. f_\alpha \right|_{x=1}$ and $\left. f_\beta \right|_{x=1}$ are \eqref{eq:f-alpha-beta} evaluated at $|\alpha|^2 = x = 1$. This quantity tells how precise a clock is needed to time the search. If measurement occurs at the wrong time, there is some probability of finding an unmarked vertex. A very small value of $\Delta t^{(0)}$ coupled with a less precise clock means the algorithm may need to be run multiple times to ensure at least one run has found a marked vertex.


\section{\label{sec:peaks}Regions of Behavior}

In this section, we will prove several behaviors that we saw in \fref{fig:prob-k3}. First, we will derive a critical value of $h$, below which the success probability peaks at 1 in a periodic fashion, and at and above which the success probability plateaus. Second, we will prove that for cases where the success probability peaks, it has sharp peaks when $h < k$ and wide peaks when $h \ge k$.


\subsection{Peaks vs Plateaus}

As shown in \cite{Wong3,Wong4}, the jumping rate $\gamma$ of the quantum walk was chosen to take the critical value \eqref{eq:gamma-critical} so that the eigenvectors of $H$ are proportional to $\ket{s} \pm \sum_{i \in M} \ket{i}/\sqrt{k}$, and so the system periodically evolves between the uniform superposition state $\ket{s}$ \eqref{eq:s} and the uniform superposition over the $k$ marked vertices $\sum_{i \in M} \ket{i}/\sqrt{k}$ \cite{Wong10C}. In this section, we will prove that this occurs when $h$ is less than some critical value $h_c$, as in \fref{fig:prob-k3-h1} through \fref{fig:prob-k3-h3.08}.

In contrast, we will prove that when $h \ge h_c$, the evolution is interrupted by $\gamma$ approaching zero, meaning the walker slows down, and the system approaches a stationary state. This can also be seen in the evolution equations in the 2D subspace, i.e., \eqref{eq:eom}, since the matrix becomes diagonal when $\gamma = 0$. This corresponds to the plateauing behavior seen in \fref{fig:prob-k3-h3.0091} through \fref{fig:prob-k3-h5}.

To show this, note from \eqref{eq:gamma-critical} that $\gamma_c$ depends on $f_\alpha$ and $f_\beta$ in \eqref{eq:f-alpha-beta}. For the general cubic-quintic nonlinearity, $f(p) = p - h p^2$, and so \eqref{eq:f-alpha-beta} with $x = |\alpha|^2$ and $|\beta|^2 = 1 - |\alpha|^2 = 1-x$ becomes
\begin{equation}\label{eq:f-alpha-beta-cubic-quintic}
\begin{aligned}
    f_\alpha &= \frac{x}{k} - h \left( \frac{x}{k} \right)^2, \\
    f_\beta &= \frac{1 - x}{N - k} - h \left( \frac{1-x}{N - k} \right)^2.
\end{aligned}
\end{equation}
Substituting \eqref{eq:f-alpha-beta-cubic-quintic} into \eqref{eq:gamma-critical}, setting it equal to zero, and simplifying, we get the following quadratic equation for the stationary values of $x$:
\begin{align*}
    \label{eq:quadratic}
    &hN(N - 2k) x^2 - \left[ kN(N-k) - 2hk^2 \right] x \\
    &\quad + k^2 (N-k-h) - \frac{k^2 (N-k)^2}{g} = 0.
\end{align*}
The solutions to this are
\begin{widetext}
\begin{equation}
    \label{eq:x-pm}
    x_\pm = \frac{kN(N-k) - 2hk^2 \pm k(N-k) \sqrt{(N-2h)^2 + \frac{4hN(N-2k)}{g}}}{2hN(N-2k)}.
\end{equation}
\end{widetext}
If one of the solutions $x_\pm$ is between $k/N$ and $1$, inclusive, the success probability will plateau. That is, if $x_\pm = k/N$, the success probability will stay at its initial value of $k/N$, and if $k/N < x_\pm \le 1$, it will rise from its initial value of $k/N$ and asymptotically approach $x_\pm$. On the other hand, if both solutions $x_\pm$ lie outside $k/N$ and $1$, the success probability oscillates between $k/N$ and $1$.

We assume that the number of marked vertices $k$ scales less than $N$, i.e., $k = o(N)$ in asymptotic notation. This is because if the number of marked vertices scales with $N$, then $k = cN$ for some constant $0 < c \le 1$. Then, it is possible to guess a marked vertex with probability $c$, and so it takes, on average, $1/c$ guesses to find a marked vertex, which is a constant runtime, eliminating the need for a quantum walk-based search algorithm. We also assume that $h$ scales less than $N$, i.e., $h = o(N)$. Then, \eqref{eq:x-pm} becomes, to leading order,
\begin{equation}
    \label{eq:x-pm-largeN}
    x_\pm = \left( 1 \pm \sqrt{\frac{g+4h}{g}} \right) \frac{k}{2h}.
\end{equation}
The second solution $x_-$ will not interrupt the evolution, since it is negative for all $g > 0$ and $h > 0$, so it does not lie between the initial success probability of $k/N$ and $1$. Then, when the first solution $x_+$ is between $k/N$ and $1$, the success probability rises from its initial value of $k/N$ and asymptotically approaches the plateau of $x_+$.

\begin{figure}
\begin{center}
    \includegraphics{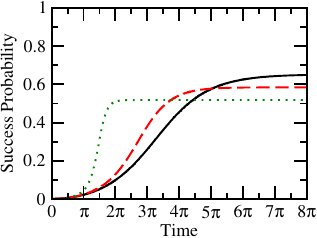}
    \caption{\label{fig:prob-n1000-gvaries}Success probability as a function of time for search on the complete graph of $N = 1000$ vertices, with $k = 2$ marked vertices, using the general cubic-quintic nonlinearity \eqref{eq:cubic-quintic} with $h = 4$. The solid black curve is $g = 10$, the dashed red curve is $g = 20$, and the dotted green curve is $g = 100$.}
\end{center}
\end{figure}

An example is shown in \fref{fig:prob-n1000-gvaries}, where we plot the success probability as a function of time for search on the complete graph of $N = 1000$ vertices with $k = 2$ marked vertices, using the general cubic-quintic nonlinearity with $h = 4$. The solid black, dashed red, and dotted green curves correspond to $g = 10$, $20$, and $100$, respectively, and using \eqref{eq:x-pm-largeN}, $x_+ = 0.653$, $0.585$, and $0.519$, respectively. We see that the curves plateau at these values. Furthermore, as expected, the runtime is faster as the nonlinear coefficient $g$ becomes greater, i.e., the plateau is approached more quickly for larger $g$, with a constant runtime achieved when $g = N-1$, as shown in \fref{fig:prob-k3}.

Now, the critical value $h_c$ is the value of $h$ that separates the regions where $x_+ \le 1$ and $x_+ > 1$. Therefore $h_c = h(x_+ = 1)$ can be found by substituting $x_+=1$ in \eqref{eq:x-pm-largeN}, resulting in
\begin{equation}
    \label{eq:h-critical}
    h_c = k \left( 1 +\frac{k}{g} \right).
\end{equation}
For example, in \fref{fig:prob-k3}, we had $k = 3$ and $g = N-1$. Then for $N = 100$, we have $h_c = 3(1+3/99) = 3.\overline{09}$, and for $N = 1000$, we have $h_c = 3(1+3/999) = 3.\overline{009}$, where the overline denotes repeating decimals. These values are consistent with \fref{fig:prob-k3}, where the solid black curve corresponding to $N = 100$ plateaus in \fref{fig:prob-k3-h3.091} and beyond, and the dashed red curve corresponding to $N = 1000$ plateaus in \fref{fig:prob-k3-h3.0091} and beyond. The height of the plateaus can be found from \eqref{eq:x-pm} or \eqref{eq:x-pm-largeN}, or when $g \gg h$ as in \fref{fig:prob-k3}, $x_+$ becomes $(1+1)k/(2h)$ to leading order, which simplifies to
\begin{equation}
    \label{eq:x-plus}
    x_+ = \frac{k}{h}.
\end{equation}
Using the values of $h$ and $k$ in \fref{fig:prob-k3-h3.0091} through \fref{fig:prob-k3-h5}, which are the graphs for which the $N = 1000$ case plateaus, we get $x_+ \approx 3/3.0091 = 0.997$, $3/3.08 = 0.974$, $3/3.091 = 0.971$, $3/3.3 = 0.909$, $3/4 = 0.75$, and $3/5 = 0.6$, in agreement with the plots. Note the slight discrepancy between the plateaus for the solid black ($N = 100$) and dashed red ($N = 1000$) curves is because \eqref{eq:x-pm-largeN} and \eqref{eq:x-plus} are for large $N$, and so the dashed red curve is in closer agreement to our calculated values.

We derived $h_c$ by examining the condition under which the jumping rate becomes  zero. There are other equivalent ways, however, of arriving at the result. One is considering \eqref{eq:runtime} for $t_*$. When its denominator $1+g(f_\alpha-f_\beta)$, evaluated at $x = 1$, is equal to zero, the runtime diverges to infinity. This corresponds to the infinite time it takes a plateau to reach a value of 1. Since this is the behavior that separates oscillatory behavior and plateaus, the value of $h$ for which $1+g(\left. f_\alpha \right|_{x=1} - \left. f_\beta \right|_{x=1}) = 0$ yields $h_c$. Comparing this with \eqref{eq:gamma-critical}, this is equivalent to the jumping rate reaching zero. Another way to derive $h_c$ is from \eqref{eq:width} for $\Delta t^{(0)}$. When its denominator $1+g(\left. f_\alpha \right|_{x=1} - \left. f_\beta \right|_{x=1}) = 0$, the width around the success probability of 1 becomes infinite, causing a plateau, and the value of $h$ corresponding to it is $h_c$.
    

\subsection{Sharp vs Wide Peaks}

Now, we consider the case when $h < h_c$, so the success probability oscillates between its initial values of \(k / N\) and 1 with a finite period. We will show that when $h < k$, the success probability peaks sharply, and when $h \ge k$, it forms broad peaks. To do this, we note that \eqref{eq:width} describes the width of the peak $\Delta t^{(0)}$ in both of these cases, since they both reach success probabilities of $1$. Evaluating \eqref{eq:f-alpha-beta-cubic-quintic} at $x = 1$ and substituting it into \eqref{eq:width}, we get
\begin{equation}
	\label{eq:width-cubic-quintic}
	\Delta t^{(0)} = \frac{2N}{1+(g/k)(1-h/k)} \sqrt{\frac{\epsilon}{k(N-k)}}.
\end{equation}

When $h < k$, $1-h/k = \Theta(1)$, and so \eqref{eq:width-cubic-quintic} is
\[ \Delta t^{(0)} = \Theta {\left( \frac{1}{1+g/k} \sqrt{\frac{N}{k}} \right)}. \]
In the denominator, whether $1$ or $g/k$ dominates depends on whether $g \ll k$ or $g \gg k$, respectively, resulting in
\begin{equation}
	\Delta t^{(0)} = \begin{cases}
        \Theta {\left( \sqrt{\frac{N}{k}} \right)}, & g \ll k, \\
        \Theta {\left( \frac{\sqrt{kN}}{g} \right)}, & g \gg k. \\
    \end{cases} \label{eq:width-sharp}
\end{equation}
In \fref{fig:prob-k3}, we had $g = N-1$ and $k = 3$, and so the second case applies, resulting in a width of
\[ \Delta t^{(0)} = \Theta {\left( \frac{1}{\sqrt{N}} \right)}. \]
Thus, we should see the width of the peak narrow as $N$ increases, and this agrees with \fref{fig:prob-h1-k2} and \fref{fig:prob-k3-h1} through \fref{fig:prob-k3-h2.99}.

Next, when $h = k$, the term $1 - h/k$ in the denominator of \eqref{eq:width-cubic-quintic} equals zero, and the dependence on the nonlinear coefficient $g$ disappears. Then, for large $N$,
\begin{equation}
    \label{eq:width-wide}
    \Delta t^{(0)} = \Theta {\left( \sqrt{\frac{N}{k}} \right)},
\end{equation}
so the peak gets wider as $N$ gets larger, which is consistent with \fref{fig:prob-h1-k1} and \fref{fig:prob-k3-h3}.

Finally, when $h > k$, examining \eqref{eq:width-cubic-quintic}, we see that as $h$ increases, the denominator gets smaller, so $\Delta t^{(0)}$ gets larger. Therefore, when $h \ge k$, the width of the peak is lower bounded by $\sqrt{N/k}$, or in asymptotic notation, $\Delta t^{(0)} = \Omega(\sqrt{N/k})$. To see just how large the width can get, consider the limit when $h = h_c$. Then, $1 - h/k = -k/g$, and the denominator of \eqref{eq:width-cubic-quintic} becomes $1 - 1 = 0$, resulting in an infinite width. This is consistent with one of our alternative ways of deriving $h_c$ as the value of $h$ for which the width is infinite. Finally, note that this region where $k < h < h_c = k(1+k/g)$ is vanishingly small for $g \gg k$, hence the small changes in $h$ going from \fref{fig:prob-k3-h3} to \fref{fig:prob-k3-h3.091}.


\section{\label{sec:runtime}Runtime}

In this section, we find the runtime of the algorithm for various behaviors. For the peaks when $h < h_c$, this corresponds to the time to reach a success probability of $1$. For the plateaus, since the time to reach $x_+$ is infinite, we take the time to half the plateau, or $x_+/2$. We consider these two cases separately below.


\subsection{Runtime for Peaks ($h < h_c$)}

Recall that \eqref{eq:time} gives the time that it takes for the success probability to reach a value of $x$. Substituting in for $f_\alpha$ and $f_\beta$ using \eqref{eq:f-alpha-beta-cubic-quintic}, \eqref{eq:time} becomes
\begin{widetext}
\[ t = \frac{Nk^2(N-k)^2}{2\sqrt{k}} \int_{k/N}^{x} \frac{1}{ax^2 + bx + c} \frac{1}{\sqrt{(1-x)(Nx-k)}} dx, \]
where
\begin{align}
    a &= -g h N (N-2k), \nonumber \\
    b &= gk (N^2-kN-2hk), \label{eq:abc} \\
    c &= -gk^2(N-k-h) + k^2(N-k)^2. \nonumber
\end{align}
This is analytically integrable, and the solution is
\begin{equation}
    \begin{split}
        \label{eq:time-to-prob}
        t = \frac{N k^2(N-k)^2}{2\sqrt{k}} \sqrt{\frac{2}{\Sigma \Delta}} &\Bigg[ 
        \frac{2 a+b+\sqrt{\Delta}}{\sqrt{\xi -\sqrt{\Delta} (N-k)}}
        \tan^{-1} \Bigg( \sqrt{\frac{2\Sigma}{\xi -\sqrt{\Delta} (N-k)}} \sqrt{\frac{Nx-k}{1-x}} \Bigg) \\
        &\quad + \frac{-2 a-b+\sqrt{\Delta}}{\sqrt{\xi +\sqrt{\Delta}(N-k)}} 
        \tan^{-1} \Bigg( \sqrt{\frac{2\Sigma}{\xi +\sqrt{\Delta} (N-k)}} \sqrt{\frac{Nx-k}{1-x}} \Bigg) 
        \Bigg],
    \end{split}
\end{equation}
where
\begin{align}
    \Delta &= b^2-4a c = k^2(N-k)^2 \left[ g^2(N-2h)^2 + 4ghN(N-2k) \right], \nonumber \\
    \Sigma &= a+b+c = (N-k)^2 \left[ k^2 + g(k-h) \right], \label{eq:DeltaSigmaxi} \\
    \xi &= 2ak + 2cN + b(N+k) = k(N-k)^2 \left[ 2Nk + g(N-2h) \right]. \nonumber
\end{align}
\end{widetext}
When $h < h_c$, the success probability peaks at 1, and so the runtime is the time it takes for the success probability to reach $1$, i.e., \eqref{eq:runtime}. Then, evaluating \eqref{eq:time-to-prob} with $x = 1$, both arctangents equal $\pi/2$, resulting in
\begin{align}
    t_*
        &= \frac{\pi}{2} \frac{N k^2(N-k)^2}{2\sqrt{k}} \sqrt{\frac{2}{\Sigma \Delta}} \nonumber \\
        &\quad\times \Bigg[ \frac{2 a+b+\sqrt{\Delta}}{\sqrt{\xi - \sqrt{\Delta}(N-k)}} + \frac{-2 a-b+\sqrt{\Delta}}{\sqrt{\xi + \sqrt{\Delta}(N-k)}} \Bigg]. \label{eq:runtime-peaks}
\end{align}
We can find the asymptotic scaling of this for large $N$ and $h \le k$, assuming as before that $h$ and $k$ both scale less than $N$. The work is shown in Appendix~\ref{appendix:general-cubicquintic}, and we get
\begin{equation}
    \label{eq:runtime-peaks-scaling}
    t_* = \begin{cases}
            \Theta(\sqrt{N/g}), & h = k, g \gg h, \\
            \Theta(\sqrt{N/k}), & h = k, g \ll h, \\
            \Theta(\sqrt{N/g}), & h < k, g \gg h, g \gg k, \\
            \Theta(\sqrt{N/k}), & h < k, g \gg h, g \ll k, \\
            \Theta(\sqrt{N/k}), & h < k, g \ll h. \\
        \end{cases}
\end{equation}
When $h > k$ (but $h < h_c$), \eqref{eq:runtime-peaks} still holds, but it is challenging to analyze because as $h$ increases, the broad peak in success probability widens more as it expands into a plateau, as seen in \fref{fig:prob-k3-h3.008} through \fref{fig:prob-k3-h3.08}, and as we analyzed in the paragraph after \eqref{eq:width-wide}. As such, the runtime goes to infinity as $h \to h_c$. Nevertheless, reaching a success probability of exactly 1 is not necessary, as having a success probability that is approximately 1 is sufficient for our analysis, and this is achieved anywhere on the top of the wide peak. Then, it is sufficient to use the runtime from the $h = k$ case in \eqref{eq:runtime-peaks-scaling} for the $h > k$ case as well, as the success probability has already jumped up at this time and will be near 1.

\begin{figure}
\begin{center}
    \includegraphics{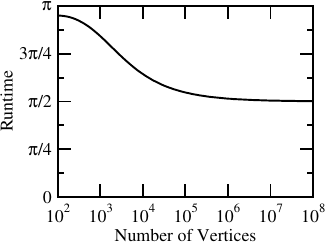}
    \caption{Plot of the runtime as a function of the number of vertices of the complete graph. The parameters are $k = 3$, $h=2.99$, and $g=N-1$. The runtime converges to $\pi/2 \approx 1.57$.}
    \label{fig:runtime-k3-h2.99}
\end{center}
\end{figure}

As a check of \eqref{eq:runtime-peaks-scaling}, the constant-runtime algorithm with $h=k$ depicted in \fref{fig:prob-k3-h3} corresponds to the first case of \eqref{eq:runtime-peaks-scaling}, since we had used $g = N-1$, and we get $t_* = \Theta(1)$, which is a constant runtime. As another example, in \fref{fig:prob-k3-h1} through \fref{fig:prob-k3-h2.99}, we have $h < k$ and $g = N-1$, so they correspond to the third case of \eqref{eq:runtime-peaks-scaling}, which yields  $t_* = \Theta(1)$. This constant time is more easily seen in \fref{fig:prob-k3-h1} and \fref{fig:prob-k3-h2}, but it is not evident in \fref{fig:prob-k3-h2.9} and \fref{fig:prob-k3-h2.99} because for these values of $N$, the plots are starting to transition to the $h = k$ wide peak. With larger $N$, however, the runtime is indeed constant, as shown in \fref{fig:runtime-k3-h2.99}, where the runtime converges to $\pi/2$ as $N$ increases.

\begin{figure*}
\begin{center}
	\includegraphics{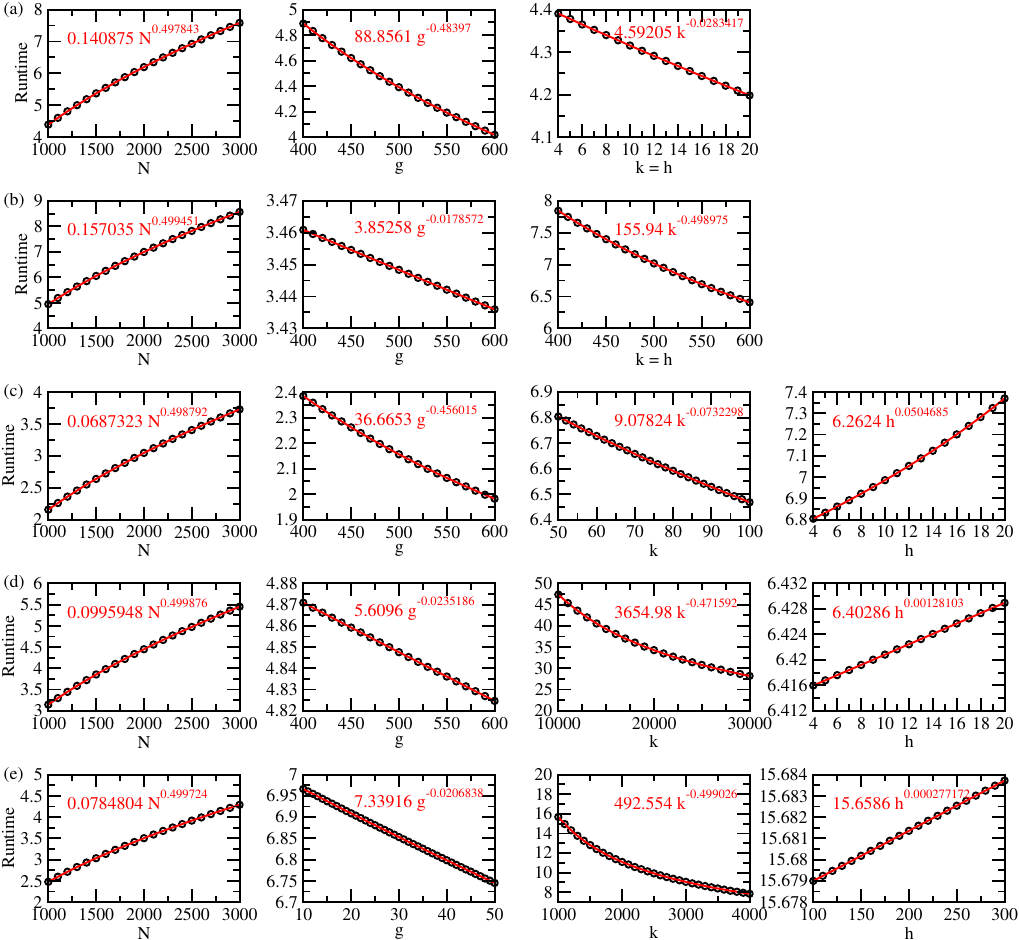}
	\caption{\label{fig:runtime-scaling} Runtime of the search algorithm with $h \le k$ with parameters from \tref{table:runtime-scaling}. In all plots, the black circles are the exact runtimes from \eqref{eq:runtime-peaks}, and the solid red curves are the best-fit power functions.}
\end{center}
\end{figure*}

As another check of \eqref{eq:runtime-peaks-scaling}, in \fref{fig:runtime-scaling}, we have plotted \eqref{eq:runtime-peaks} for each of the five cases in \eqref{eq:runtime-peaks-scaling}. Subfigure (a)'s three graphs correspond to the first case, subfigure (b)'s three graphs correspond to the second case, and so forth. The black circles are the actual runtime values given by \eqref{eq:runtime-peaks}, and the solid red curves are power function fits. In the three plots in subfigure (a), the first plot indicates that the runtime scales as $\sqrt{N}$, the second that it scales as $1/\sqrt{g}$, and the third that it does not depend on $h = k$. Altogether, these indicate that the runtime scales as $\sqrt{N/g}$, in agreement with the first case of \eqref{eq:runtime-peaks-scaling}. Similarly, subfigures (b) through (e) are consistent with the second through fifth cases of \eqref{eq:runtime-peaks-scaling}.

\begin{table*}
    \caption{\label{table:runtime-scaling} Parameters for the plots in \fref{fig:runtime-scaling}.}
    \begin{ruledtabular}
    \begin{tabular}{c|c|c|c|c}
        Subfigure & Plot 1 & Plot 2 & Plot 3 & Plot 4 \\
        \colrule
        (a)       & $N = 1000, 1100, \dots, 3000$ & $N = 1000$                 & $N = 1000$                & \\
        $h = k$   & $g = 500$                     & $g = 400, 410, \dots, 600$ & $g = 500$                 & Not Applicable \\
        $g \gg h$ & $h = k = 4$                   & $h = k = 4$                & $h = k = 4, 5, \dots, 20$ & \\
        \colrule
        (b)       & $N = 1000, 1100, \dots, 3000$ & $N = 10000$                & $N = 10000$                    & \\
        $h = k$   & $g = 4$                       & $g = 400, 410, \dots, 600$ & $g = 4$                        & Not Applicable \\
        $g \ll h$ & $h = k = 100$                 & $h = k = 2000$             & $h = k = 400, 410, \dots, 600$ & \\
        \colrule
        (c)       & $N = 1000, 1100, \dots, 3000$ & $N = 1000$                 & $N = 10000$              & $N = 10000$ \\
        $h < k$   & $g = 500$                     & $g = 400, 410, \dots, 600$ & $g = 500$                & $g = 500$ \\
        $g \gg h$ & $k = 50$                      & $k = 50$                   & $k = 50, 52, \dots, 100$ & $k = 50$ \\
        $g \gg k$ & $h = 4$                       & $h = 4$                    & $h = 4$                  & $h = 4, 5, \dots, 20$ \\
        \colrule
        (d)       & $N = 1000, 1100, \dots, 3000$ & $N = 100000$               & $N = 10000000$                   & $N = 10000$ \\
        $h < k$   & $g = 50$                      & $g = 400, 410, \dots, 600$ & $g = 1000$                       & $g = 100$ \\
        $g \gg h$ & $k = 200$                     & $k = 10000$                & $k = 10000, 11000, \dots, 30000$ & $k = 500$ \\
        $g \ll k$ & $h = 4$                       & $h = 4$                    & $h = 4$                          & $h = 4, 5, \dots, 20$ \\
        \colrule
        (e)       & $N = 1000, 1100, \dots, 3000$ & $N = 10000$             & $N = 100000$                  & $N = 100000$ \\
        $h < k$   & $g = 4$                       & $g = 10, 11, \dots, 50$ & $g = 4$                       & $g = 4$ \\
        $g \ll h$ & $k = 400$                     & $k = 500$               & $k = 1000, 1100, \dots, 4000$ & $k = 1000$ \\
                  & $h = 200$                     & $h = 100$               & $h = 100$                     & $h = 100, 110, \dots, 300$ \\
    \end{tabular}
    \end{ruledtabular}
\end{table*}

When $g = N-1$, we can calculate the precise runtimes of the peaks that we saw in \fref{fig:prob-k3-h1}. Beginning with $h < k$, the terms in brackets in \eqref{eq:runtime-peaks} are asymptotically
\[ \frac{2 a+b+\sqrt{\Delta}}{\sqrt{\xi -\sqrt{\Delta} (N-k)}} \approx \sqrt{\frac{2(k-h)}{k}} N^{3/2}, \]
and
\[ \frac{-2 a-b+\sqrt{\Delta}}{\sqrt{\xi +\sqrt{\Delta}(N-k)}} \approx \frac{\sqrt{2}hN}{\sqrt{k}}. \]
Meanwhile, the overall factors in \eqref{eq:runtime-peaks} become
\[ \frac{\pi}{2} \frac{N k^2(N-k)^2}{2\sqrt{k}} \sqrt{\frac{2}{\Sigma \Delta}} \approx \frac{\pi}{2} \sqrt{\frac{k}{2(k-h)}} \frac{1}{N^{3/2}}. \]
Putting all these together in \eqref{eq:runtime-peaks}, for large $N$, the time it takes to reach a success probability of $1$ is
\[ t_* = \frac{\pi}{2} \sqrt{\frac{k}{2(k-h)}} \frac{1}{N^{3/2}} \left( \sqrt{\frac{2(k-h)}{k}} N^{3/2} + \frac{\sqrt{2}hN}{\sqrt{k}} \right). \]
For large $N$, the first term in parenthesis dominates, so we get
\begin{equation}
    \label{eq:runtime-sharp-constant}
     t_* = \frac{\pi}{2},
\end{equation}
which is consistent with \fref{fig:prob-k3-h1} through \fref{fig:prob-k3-h2.99} and \fref{fig:runtime-k3-h2.99}.

Next, when $h = k$, the terms in brackets in \eqref{eq:runtime-peaks} are asymptotically
\[ \frac{2 a+b+\sqrt{\Delta}}{\sqrt{\xi -\sqrt{\Delta} (N-k)}} \approx \sqrt{2gkN}, \]
and
\[ \frac{-2 a-b+\sqrt{\Delta}}{\sqrt{\xi +\sqrt{\Delta}(N-k)}} \approx \sqrt{2gkN}. \]
That is, they are asymptotically the same. The overall factors in \eqref{eq:runtime-peaks} become
\[ \frac{\pi}{2} \frac{N k^2(N-k)^2}{2\sqrt{k}} \sqrt{\frac{2}{\Sigma \Delta}} \approx \frac{\pi}{2\sqrt{2}g\sqrt{k(1+4k/g)}}. \]
Putting all these together in \eqref{eq:runtime-peaks}, for large $N$, the time it takes to reach a success probability of $1$ is
\begin{equation}
    \label{eq:runtime-wide}
    t_* = \pi\sqrt{\frac{N}{g(1+4k/g)}}.
\end{equation}
Increasing either the overall nonlinear coefficient ($g$), or relative strength of the quintic nonlinearity and number of marked vertices ($h = k$) tends to decrease the runtime. For $g = \Theta(N)$, this gives \begin{equation}
    \label{eq:runtime-wide-constant}
     t_* = \pi,
\end{equation}
which is consistent with \fref{fig:prob-h1-k1} and \fref{fig:prob-k3-h3}.


\subsection{Time to Plateau ($h \ge h_c$)}

Now, we will find the runtime when the success probability plateaus in the limits that $N$ is large compared to both $h$ and $k$, and $g$ is large compared to $k$, such as when $g = N-1$. The large $g$ limit is helpful to simplify the arctangents, but other cases can be done numerically. We can again use \eqref{eq:time-to-prob} to find the time it takes the success probability to reach a value of $x$. From \eqref{eq:x-plus}, when $h \ge h_c$, , the success probability plateaus toward $x_+ = k/h$. Then, we might set $x = k/h$ in \eqref{eq:time-to-prob} to find the runtime, but this will result in an infinite runtime since $x = k/h$ is an asymptote of the success probability, and so it takes infinite time to reach it. To explicitly show this, when $x = k/h$, the argument of the second arctangent in \eqref{eq:time-to-prob} becomes, for large $N$ (and $g = N-1)$,
\begin{align*}
    \sqrt{\frac{2\Sigma}{\xi + \sqrt{\Delta} (N-k)}} \sqrt{\frac{Nx - k}{1 - x}} 
    &\approx \sqrt{\frac{-(h-k)}{kN}} \sqrt{\frac{N k/h}{1 - k/h}} \\ 
    &= \sqrt{-1} = i.
\end{align*}
Then, the second arctangent in \eqref{eq:time-to-prob} is
\[\tan^{-1}(i) = \infty, \]
which results in an infinite runtime $t = \infty$.

Instead, we take the runtime to be the time it takes to reach half the maximum success probability, i.e., $x = x_+ / 2 \approx k / 2h$. In \fref{fig:prob-k3-h3.0091} through \ref{fig:prob-k3-h5}, we see that the time to reach $x = x_+ / 2$ is a good marker for when the success probability jumps up to the plateau, since the rise is very sharp. Taking large $N$ (and $g = N-1$) as well as $x = k / 2h$, the first arctangent in \eqref{eq:time-to-prob} is
\[ \tan^{-1} \left( \sqrt{\frac{1}{k}}\sqrt{\frac{kN}{2h-k}} \right) \approx \frac{\pi}{2}, \]
and the second arctangent is
\begin{align*}
    &\tan^{-1} \left( \sqrt{\frac{-(h-k)}{kN}} \sqrt{\frac{kN}{2h - k}} \right) \\
    &\quad = \tan^{-1} \left( \sqrt{\frac{-(h-k)}{2h - k}} \right) \\
    &\quad = \frac{i}{2}
        \ln \left[ 
        \frac{\sqrt{2h-k} + \sqrt{h-k}}{\sqrt{2h-k} - \sqrt{h-k}}\
        \right],
\end{align*}
where in the last line, we used the identity
\begin{equation*}
    \tan^{-1}(ix) = \frac{i}{2} \ln \left( \frac{1 + x}{1 - x} \right).
\end{equation*}
The factors in front of the arctangents are
\[ \frac{2 a+b+\sqrt{\Delta}}{\sqrt{\xi -\sqrt{\Delta} (N-k)}} \approx \sqrt{\frac{-2(h-k)}{k}} N^{3/2}, \]
and
\[ \frac{-2 a-b+\sqrt{\Delta}}{\sqrt{\xi +\sqrt{\Delta}(N-k)}} \approx \frac{\sqrt{2}hN}{\sqrt{k}}. \]
Finally, the overall factors in \eqref{eq:time-to-prob} become
\[ \frac{N k^2(N-k)^2}{2\sqrt{k}} \sqrt{\frac{2}{\Sigma \Delta}} \approx \sqrt{\frac{k}{-2(h-k)}} \frac{1}{N^{3/2}}. \]
Substituting all of these into \eqref{eq:time-to-prob}, the time it takes to reach a success probability of $x_+/2 = k/2h$, for large $N$, is
\begin{align}
    t
        &= \sqrt{\frac{k}{-2(h-k)}} \frac{1}{N^{3/2}} \Bigg[ \sqrt{\frac{-2(h-k)}{k}} N^{3/2} \frac{\pi}{2} \nonumber \\
        &\quad\quad\quad\quad\quad\quad + \frac{\sqrt{2}hN}{\sqrt{k}} \frac{i}{2} \ln \bigg( \frac{\sqrt{2h-k} + \sqrt{h-k}}{\sqrt{2h-k} - \sqrt{h-k}}\bigg) \Bigg] \nonumber \\
        &= \frac{\pi}{2} + \frac{h}{2\sqrt{(h-k)N}} \ln \bigg( \frac{\sqrt{2h-k} + \sqrt{h-k}}{\sqrt{2h-k} - \sqrt{h-k}}\bigg) \nonumber \\
        &\approx \frac{\pi}{2}. \label{eq:runtime-plateau-constant}
\end{align}
This agrees with \fref{fig:prob-k3-h3.0091} through \fref{fig:prob-k3-h5}, where the success probability cliff of the plateau occurs at $t = \pi/2$.


\section{\label{sec:resources}Additional Resources}

As discussed in the introduction, Zalka's lower bound on quantum search with parallel oracles \eqref{eq:Zalka} implies that any speedup over Grover's quadratic runtime comes at the expense of the number of particles needed in the many-body quantum system. In this section, we quantify this by exploring the lower-bound on the number of atoms needed in the Bose-Einstein condensate evolving with the cubic-quintic nonlinearity. We also saw that when $h < k$, the success probability can peak sharply, necessitating a sufficiently precise clock, and we quantify this clock resource as well.

\subsection{\label{sec:many-body-resources}Many-Body Resources}

In our many-body quantum system undertaking search by quantum walk, the number of particles is equivalent to a number of parallel oracles. Then, Zalka's result \eqref{eq:Zalka} puts a lower bound on the number of particles needed for the many-body quantum system to search more quickly than $O(\sqrt{N})$ time.

For a Bose-Einstein condensate, this would be the number of condensate atoms $n_\text{BEC}$. Then, \eqref{eq:Zalka} implies that $n_\text{BEC} t_*^2 = \Omega(N)$ \cite{Zalka1999}, and so
\begin{equation}
    \label{eq:Zalka-BEC}
     n_\text{BEC} = \Omega {\left( \frac{N}{t_*^2} \right)}.
\end{equation}
Since the choice of $g$ and $h$ affects the runtime of the algorithm, we get different lower bounds on the number of condensate atoms depending on the choice of parameters. For example, the constant-runtime algorithm takes $g = N-1$ and $h \ge k$, and with these parameters, we need at least $\Omega(N)$ condensate atoms. As another example, with $g = \Theta(\sqrt{N})$ and arbitrary $h$, the runtime is $O(N^{1/4})$, and these parameters would only require at least $\Omega(\sqrt{N})$ atoms. As still another example, when $g = 0$, we have the linear algorithm, which searches in $O(\sqrt{N})$ time, and the number of condensate atoms is $\Omega(1)$, which makes sense as there is no need for many condensate atoms to interact and give rise to an effective nonlinearity.

Since our results depend on the many-body quantum system evolving with the cubic-quintic nonlinearity, these results can also be interpreted as lower bounds on the number of condensate atoms needed for a Bose-Einstein condensate to effectively evolve by the nonlinearity. In particular cases, there could be additional resources beyond the minimum ones identified here.

Note that previous results on searching with other nonlinearities \cite{Wong3,Wong4} focused on the lower bound that can be obtained from a constant-runtime algorithm, whereas our argument is that different lower bounds exist for different speed algorithms with different parameters.

\subsection{\label{sec:clock-resource}Clock Resources}

In the case when $h < k$, we showed in \eqref{eq:width-sharp} that the sharp peaks in the success probability decrease in width with increasing $N$ if $g=N-1$, while from  \eqref{eq:runtime-sharp-constant}, the runtime stays constant. Achieving this runtime therefore requires extra resources from the clock needed to achieve a sufficiently high time-measurement precision to stop at the peak in the success probability, as in \fref{fig:prob-k3-h1} and \fref{fig:prob-k3-h2}.

As shown in \eqref{eq:width-sharp}, the peak has a width of $O(1/\sqrt{N})$. An atomic clock that utilizes entanglement between $n_\text{clock}$ oscillators has a theoretical precision that scales with $1/(n_\text{clock}\sqrt{\tau})$, where $\tau$ is the measurement time. Considering constant $\tau$, $n_\text{clock}=O(\sqrt{N})$ clock atoms are needed to achieve the required scaling of the time-measurement precision. Likening this to a ``space'' requirement, the overall ``space $\times$ time'' resource for the constant-runtime algorithm is $O(\sqrt{N})$, i.e., the same as the linear algorithm that takes $O(\sqrt{N})$ time and $O(1)$ clock ions.

We can reduce this resource by choosing $g$ such that the time-measurement resources are less at the expense of a slower runtime. Taking the second case of \eqref{eq:width-sharp} with $k$ constant, $\Delta t^{(0)} = \Theta(\sqrt{N}/g)$. When $g \ll \sqrt{N}$, this implies an unphysical decreased scaling in the number of clock ions needed to perform the measurement. Therefore, the clock resource requirement is $O[\max(1/\Delta t^{(0)}, 1)]$, which scales at least as  a constant. The space/clock resource is therefore
\[ \text{``space''} = \begin{cases}
    O(1), & g \ll \sqrt{N}, \\
    O(g/\sqrt{N}), & g \gg \sqrt{N}. \\
\end{cases} \]
For the runtime, we take the third case of \eqref{eq:runtime-peaks-scaling}, $t_* = \Theta(\sqrt{N/g})$. Multiplying this by the space/clock resource, the overall ``space $\times$ time'' resource is
\[ \text{``space $\times$ time''} = \begin{cases}
    O(\sqrt{N/g}), & g \ll \sqrt{N}, \\
    O(\sqrt{g}), & g \gg \sqrt{N}. \\
\end{cases} \]
This is minimized at the boundary, i.e., $g = \Theta(\sqrt{N})$, at which $\text{``space $\times$ time''} = O(N^{1/4})$, corresponding to a constant number of clock ions and a runtime of $O(N^{1/4})$.

For $h = k$, the width is wide, so time-measurement precision is not a concern. In particular, we showed in \eqref{eq:width-wide} that the width scales as $\Theta(\sqrt{N/k})$, and so the clock resource scales as a constant. Then, the overall ``space $\times$ time'' resource is just the runtime, which is asymptotically a constant when taking $g = \Theta(N)$ \eqref{eq:runtime-wide}.

For $h > k$ but $h < h_c$, the width increases more, so the time-measurement precision is again not a concern, and a constant ``space $\times$ time'' resource is still obtainable.

Similarly, when $h \ge h_c$, the plateau indicates that constant time-measurement precision is sufficient to catch the success probability near its maximum. Then, as with wide peaks, the overall ``space $\times$ time'' resource is just the runtime, which for $g = \Theta(N)$ is a constant \eqref{eq:runtime-plateau-constant}.

Altogether, when $h \ge k$, a constant ``space $\times$ time'' algorithm is attainable, and we can tune $g$ and $h$ to select this regime.


\section{\label{sec:conclusion}Conclusion}

In this paper, we explored the speedup that can be obtained for the unstructured search problem using a many-body quantum system that queries the oracle in parallel. We focused on systems that effectively evolve by a general cubic-quintic nonlinear Schr\"odinger equation, where the nonlinear coefficient was $g$ and the ratio of the coefficients of the quintic and cubic terms was $h$, conducting a continuous-time quantum walk on the complete graph of $N$ vertices to search for one of $k$ marked vertices. We proved that constant-runtime algorithms were obtained when $g = N-1$, but with different features depending on the parameters. When $h < h_c$, the success probability reaches 1, and the evolution is periodic. This oscillatory behavior can be further divided into cases where $h < k$ and $h \ge k$, respectively corresponding to sharp and wide peaks. When $h \ge h_c$, the success probability asymptotically approaches a constant value, so it may be necessary to repeat the algorithm an expected constant number of times, which does not affect the overall scaling of the runtime.

The case of sharp peaks ($h < k$) requires high time-measurement precision, resulting in additional resources for a sufficiently precise clock. This can be alleviated by opting for a slower runtime, such as $O(N^{1/4})$, or by tuning $h \ge k$ so that the success probability forms a wide peak or plateau. In all cases, any speedup over $O(\sqrt{N})$ comes at the expense of additional particles in the many-body quantum system, and this also gives a lower bound on the number of particles needed for the system to evolve by the effective nonlinearity.


\begin{acknowledgments}
    This material is based upon work supported in part by the National Science Foundation EPSCoR Cooperative Agreement OIA-2044049, Nebraska’s EQUATE collaboration. Any opinions, findings, and conclusions or recommendations expressed in this material are those of the author(s) and do not necessarily reflect the views of the National Science Foundation.
\end{acknowledgments}


\begin{appendix}

\section{\label{appendix:general-cubicquintic}Scaling of Runtime Terms when $h \le k$}

In this Appendix, we derive \eqref{eq:runtime-peaks-scaling}, which is the runtime of the search algorithm when $h \le k$, assuming $h$ and $k$ both scale less than $N$, i.e., $h = o(N)$ and $k = o(N)$ in asymptotic notation. We do this by calculating the asymptotic scaling of \eqref{eq:runtime-peaks} with the above assumptions.

We begin with the two fractions in brackets in \eqref{eq:runtime-peaks}, starting with the numerator of each fraction. Using \eqref{eq:abc} and \eqref{eq:DeltaSigmaxi}, we get
\begin{align*}
    2a&+b+\sqrt{\Delta} \\
        &= g(N-k)(kN-2h(N-k)) \\
        &\quad+ k(N-k) \sqrt{g^2(N-2h)^2 + 4ghN(N-2k)} \\
        &\approx gN(kN-2hN) + kN \sqrt{g^2N^2 + 4ghN^2} \\
        &= gkN^2 - 2ghN^2 + kN^2 \sqrt{g^2 + 4gh}.
\end{align*}
Note throughout this appendix, we use $\approx$ to denote ``of the same order as,'' permitting us to only keep dominant terms and drop constant factors. Then, our use of $\approx$ is equivalent to Knuth's big-$\Theta$ in asymptotic notation, but it prevents clutter from $\Theta$'s appearing in each subsequent expression. Now, depending on whether $g$ or $h$ dominate each other, we get different behaviors for the square root. When $g \gg h$, we Taylor expand the square root to second order, i.e., $\sqrt{g^2+4gh} \approx g + 2h$, and when $g \ll h$, we take $\sqrt{g^2+4gh} \approx \sqrt{4gh}$:
\begin{align*}
    2a&+b+\sqrt{\Delta} \\
        &\approx \begin{cases}
            gkN^2 - 2ghN^2 + kN^2 (g + 2h), & g \gg h, \\
            gkN^2 - 2ghN^2 + kN^2 \sqrt{4gh}, & g \ll h, \\
        \end{cases} \\
        &= \begin{cases}
            2g(k-h)N^2 + 2hkN^2, & g \gg h, \\
            gkN^2 - 2ghN^2 + kN^2 \sqrt{4gh}, & g \ll h. \\
        \end{cases}
\end{align*}
In the first case, which term dominates depends on whether $h = k$ or $h < k$. In the second case, the last term always dominates, since $h \le k$ and $g \ll h$. Then,
\begin{align}
    2a+b+\sqrt{\Delta} 
        &\approx \begin{cases}
            2hkN^2, & g \gg h, h = k, \\
            2gkN^2, & g \gg h, h < k, \\
            kN^2 \sqrt{4gh}, & g \ll h, \\
        \end{cases} \nonumber \\
        &\approx \begin{cases}
            k^2N^2, & g \gg h, h = k, \\
            gkN^2, & g \gg h, h < k, \\
            kN^2\sqrt{gh}, & g \ll h, \\
        \end{cases} \label{eq:abDelta}
\end{align}
where in the last step, we dropped constant factors.

For the second numerator, we similarly get
\[ -2a-b+\sqrt{\Delta} \approx -gkN^2 + 2ghN^2 + kN^2 \sqrt{g^2 + 4gh}. \]
As before, we approximate $\sqrt{g^2 + 4gh} \approx g + 2h$ or $\sqrt{4gh}$ depending on whether $g \gg h$ or $g \ll h$, respectively, yielding
\begin{align*}
    -2a&-b+\sqrt{\Delta} \\
        &\approx \begin{cases}
            -gkN^2 + 2ghN^2 + kN^2 (g + 2h), & g \gg h, \\
            -gkN^2 + 2ghN^2 + kN^2 \sqrt{4gh}, & g \ll h, \\
        \end{cases} \\
        &\approx \begin{cases}
            2ghN^2 + 2hkN^2, & g \gg h, \\
            kN^2 \sqrt{4gh}, & g \ll h. \\
        \end{cases}
\end{align*}
In the first case, the first and second terms dominate when $g \gg k$ and $g \ll h$, respectively, so we get
\begin{equation}
    -2a-b+\sqrt{\Delta}
        \approx \begin{cases}
            ghN^2, & g \gg h, g \gg k, \\
            hkN^2, & g \gg h, g \ll k, \\
            kN^2\sqrt{gh}, & g \ll h, \\
        \end{cases} \label{eq:minusabDelta}
\end{equation}
where we also dropped constant factors.

Next, consider the denominator of each fraction. Using \eqref{eq:DeltaSigmaxi},
\begin{align*}
    \xi &- \sqrt{\Delta}(N-k) \\
        &= k(N-k)^2 \left[ 2Nk + g(N-2h) \right] \\
        &\quad- k(N-k)^2 \sqrt{g^2(N-2h)^2 + 4ghN(N-2k)} \\
        &= k(N-k)^2 \Big[ 2Nk + g(N-2h) \\
        &\quad\quad\quad\quad\quad\quad- \sqrt{g^2(N-2h)^2 + 4ghN(N-2k)} \Big] \\
        &\approx kN^2 \left[ 2Nk + gN - \sqrt{g^2N^2 + 4ghN^2} \right] \\
        &= kN^3 \left[ 2k + g - \sqrt{g^2 + 4gh} \right].
\end{align*}    
When $g \gg h$, we now Taylor expand the square root to third order, i.e., $\sqrt{g^2 + 4gh} \approx g + 2h - 2h^2/g$, and when $g \ll h$, we simply use $\sqrt{4gh}$. This yields
\begin{align*}
    \xi &- \sqrt{\Delta}(N-k) \\
        &\approx \begin{cases}
            kN^3 \left[ 2k + g - (g+2h - 2h^2/g) \right], & g \gg h, \\
            kN^3 \left[ 2k + g - \sqrt{4gh} \right], & g \ll h, \\
        \end{cases} \\
        &\approx \begin{cases}
            kN^3 \left[ 2(k-h) + 2h^2/g \right], & g \gg h, \\
            2k^2N^3, & g \ll h. \\
        \end{cases}
\end{align*}
The first case depends on whether $h = k$ or $h < k$:
\begin{align}
    \xi &- \sqrt{\Delta}(N-k) \nonumber \\
        &\approx \begin{cases}
            kN^3 \left[ 2h^2/g \right], & g \gg h, h = k, \\
            kN^3 \left[ 2k \right], & g \gg h, h < k, \\
            2k^2N^3, & g \ll h, \\
        \end{cases} \nonumber \\
        &\approx \begin{cases}
            k^3N^3/g, & g \gg h, h = k, \\
            k^2N^3, & g \gg h, h < k, \\
            k^2N^3, & g \ll h, \\
        \end{cases} \label{eq:ximinus}
\end{align}
where we also dropped constant factors.

For the other denominator, we similarly have
\[ \xi + \sqrt{\Delta}(N-k) \approx kN^3 \left[ 2k + g + \sqrt{g^2 + 4gh} \right]. \]
Now, when $g \gg h$, it suffices to Taylor expand the square root to first order, i.e., $\sqrt{g^2 + 4gh} \approx g$:
\begin{align*}
    \xi &+ \sqrt{\Delta}(N-k) \\
        &\approx \begin{cases}
            kN^3 \left[ 2k + g + g \right], & g \gg h, \\
            kN^3 \left[ 2k + g + \sqrt{4gh} \right], & g \ll h, \\
        \end{cases} \\
        &= \begin{cases}
            kN^3 \left[ 2k + 2g \right], & g \gg h, \\
            kN^3 \left[ 2k + g + \sqrt{4gh} \right], & g \ll h. \\
        \end{cases}
\end{align*}
In the first case, which term dominates depends on the relationship between $g$ and $k$, and in the second case $2k$ dominates the other terms since $h \le k$ and $g \ll h$:
\begin{align}
    \xi &+ \sqrt{\Delta}(N-k) \nonumber \\
        &\approx \begin{cases}
            kN^3 \left[ 2g \right], & g \gg h, g \gg k, \\
            kN^3 \left[ 2k \right], & g \gg h, g \ll k, \\
            kN^3 \left[ 2k \right], & g \ll h, \\
        \end{cases} \nonumber \\
        &\approx \begin{cases}
            gkN^3, & g \gg h, g \gg k, \\
            k^2N^3, & g \gg h, g \ll k, \\
            k^2N^3, & g \ll h, \\
        \end{cases} \label{eq:xiplus}
\end{align}
where we also dropped constant factors.

Now, to find the fractions in brackets in \eqref{eq:runtime-peaks}, we divide \eqref{eq:abDelta} by the square root of \eqref{eq:ximinus}, and \eqref{eq:minusabDelta} by the square root of \eqref{eq:xiplus}, resulting in
\begin{equation}
    \frac{2a+b+\sqrt{\Delta}}{\sqrt{\xi - \sqrt{\Delta}(N-k)}}
        \approx \begin{cases}
            \sqrt{gkN}, & g \gg h, h = k, \\
            g\sqrt{N}, & g \gg h, h < k, \\
            \sqrt{ghN}, & g \ll h. \\
        \end{cases} \label{eq:abDeltaxi} 
\end{equation}
and
\begin{equation}
    \frac{-2a-b+\sqrt{\Delta}}{\sqrt{\xi + \sqrt{\Delta}(N-k)}} 
        \approx \begin{cases}
            h\sqrt{gN/k}, & g \gg h, g \gg k, \\
            h\sqrt{N}, & g \gg h, g \ll k, \\
            \sqrt{ghN}, & g \ll h. \\
        \end{cases} \label{eq:abDeltaminusxi}
\end{equation}
Adding these fractions,
\begin{align*}
    &\frac{2a+b+\sqrt{\Delta}}{\sqrt{\xi - \sqrt{\Delta}(N-k)}} + \frac{-2a-b+\sqrt{\Delta}}{\sqrt{\xi + \sqrt{\Delta}(N-k)}} \\
        &\quad\approx \begin{cases}
            \sqrt{gkN} + h\sqrt{gN/k}, & g \gg h, h = k, g \gg k, \\
            \sqrt{gkN} + h\sqrt{N}, & g \gg h, h = k, g \ll k, \\
            g\sqrt{N} + h\sqrt{gN/k}, & g \gg h, h < k, g \gg k, \\
            g\sqrt{N} + h\sqrt{N}, & g \gg h, h < k, g \ll k, \\
            \sqrt{ghN} + \sqrt{ghN}, & g \ll h. \\
        \end{cases}
\end{align*}
In the first case, the terms have identical scalings since $h = k$. The second case is not possible, since $g$ cannot scale both greater and less than $h = k$. The third and fourth cases are both dominated by $g\sqrt{2N}$, and these two cases can be combined. In the fifth case, both terms have identical scalings. Then,
\begin{align}
    &\frac{2a+b+\sqrt{\Delta}}{\sqrt{\xi - \sqrt{\Delta}(N-k)}} + \frac{-2a-b+\sqrt{\Delta}}{\sqrt{\xi + \sqrt{\Delta}(N-k)}} \nonumber \\
        &\quad\approx \begin{cases}
            \sqrt{gkN}, & g \gg h, h = k, \\
            g\sqrt{N}, & g \gg h, h < k, \\
            \sqrt{ghN}, & g \ll h. \\
        \end{cases} \label{eq:runtime-peaks-fractions}
\end{align}
Ignoring constant factors, this scales identically to the first fraction's scaling in \eqref{eq:abDeltaxi}, and so the first fraction \eqref{eq:abDeltaxi} is at least as dominant as the second fraction \eqref{eq:abDeltaminusxi} in all cases.

Examining \eqref{eq:runtime-peaks} for the runtime, we now have the term in brackets, i.e., the sum of the fractions in \eqref{eq:runtime-peaks-fractions}. Now, we consider the overall factors. To determine $\Delta \Sigma$, we begin with \eqref{eq:DeltaSigmaxi}:
\begin{align*}
    \Sigma \Delta
        &= (N-k)^2 \left[ k^2 + g(k-h) \right] \\
        &\quad \times k^2(N-k)^2 \left[ g^2(N-2h)^2 + 4ghN(N-2k) \right] \\
        &\approx k^2N^4 \left[ k^2 + g(k-h) \right] \left( g^2N^2 + 4ghN^2 \right) \\
        &= gk^2N^6 \left[ k^2 + g(k-h) \right] \left( g + 4h \right).
\end{align*}
This depends on whether $h = k$ or $k \gg h$:
\begin{align*}
    \Sigma \Delta
        &\approx \begin{cases}
            gh^2N^6 [h^2] \left( g + 4h \right), & h = k, \\
            gk^2N^6 \left[ k^2 + g(k-h) \right] \left( g + 4h \right), & h < k, \\
        \end{cases} \\
        &\approx \begin{cases}
            gk^4N^6 \left( g + 4k \right), & h = k, \\
            gk^2N^6 \left[ k^2 + gk \right] \left( g + 4h \right), & h < k, \\
        \end{cases} \\
        &= \begin{cases}
            gk^4N^6 \left( g + 4k \right), & h = k, \\
            gk^3N^6 \left( k + g \right) \left( g + 4h \right), & h < k. \\
        \end{cases}
\end{align*}
Both of these cases depend on the relationship between $g$ and $h$:
\begin{align*}
    \Sigma \Delta
        &\approx \begin{cases}
            gk^4N^6 (g), & h = k, g \gg h, \\
            gk^4N^6 (4k), & h = k, g \ll h, \\
            gk^3N^6 (k+g) (g), & h < k, g \gg h, \\
            gk^3N^6 (k+g) (4h), & h < k, g \ll h, \\
        \end{cases} \\
        &= \begin{cases}
            g^2k^4N^6, & h = k, g \gg h, \\
            4gk^5N^6, & h = k, g \ll h, \\
            g^2k^3N^6 (k+g), & h < k, g \gg h, \\
            4ghk^3N^6 (k+g), & h < k, g \ll h. \\
        \end{cases}
\end{align*}
The third case depends on the relationship between $k$ and $g$, while in the fourth case, since $h < k$ and $g \ll h$, it must be that $g \ll k$, and so $k+g$ is dominated by $k$:
\begin{align}
    \Sigma \Delta \nonumber
        &\approx \begin{cases}
            g^2k^4N^6, & h = k, g \gg h, \\
            4gk^5N^6, & h = k, g \ll h, \\
            g^2k^3N^6 (g), & h < k, g \gg h, g \gg k, \\
            g^2k^3N^6 (k), & h < k, g \gg h, g \ll k, \\
            4ghk^3N^6 (k), & h < k, g \ll h, \\
        \end{cases} \nonumber \\
        &\approx \begin{cases}
            g^2k^4N^6, & h = k, g \gg h, \\
            gk^5N^6, & h = k, g \ll h, \\
            g^3k^3N^6, & h < k, g \gg h, g \gg k, \\
            g^2k^4N^6, & h < k, g \gg h, g \ll k, \\
            ghk^4N^6, & h < k, g \ll h, \\
        \end{cases} \label{eq:EpsilonDelta}
\end{align}
where we also dropped constant factors.

Finally, from \eqref{eq:runtime-peaks}, runtime scales as
\begin{align*}
    t_*
        &\approx \frac{k^{3/2}N^3}{\sqrt{\Sigma \Delta}} \Bigg[ \frac{2 a+b+\sqrt{\Delta}}{\sqrt{\xi - \sqrt{\Delta}(N-k)} } + \frac{-2 a-b+\sqrt{\Delta}}{\sqrt{\xi + \sqrt{\Delta}(N-k)}} \Bigg].
\end{align*}
Substituting \eqref{eq:runtime-peaks-fractions} and \eqref{eq:EpsilonDelta} into this, we get
\begin{align*}
    t_*
        &\approx \begin{cases}
            \frac{k^{3/2}N^3}{\sqrt{g^2k^4N^6}} \sqrt{gkN}, & h = k, g \gg h, \\
            \frac{k^{3/2}N^3}{\sqrt{gk^5N^6}} \sqrt{gkN}, & h = k, g \ll h, \\
            \frac{k^{3/2}N^3}{\sqrt{g^3k^3N^6}} g\sqrt{N}, & h < k, g \gg h, g \gg k, \\
            \frac{k^{3/2}N^3}{\sqrt{g^2k^4N^6}} g\sqrt{N}, & h < k, g \gg h, g \ll k, \\
            \frac{k^{3/2}N^3}{\sqrt{ghk^4N^6}} \sqrt{ghN}, & h < k, g \ll h. \\
        \end{cases} \\
        &= \begin{cases}
            \sqrt{N/g}, & h = k, g \gg h, \\
            \sqrt{N/k}, & h = k, g \ll h, \\
            \sqrt{N/g}, & h < k, g \gg h, g \gg k, \\
            \sqrt{N/k}, & h < k, g \gg h, g \ll k, \\
            \sqrt{N/k}, & h < k, g \ll h, \\
        \end{cases}
\end{align*}
as reported in the main text as \eqref{eq:runtime-peaks-scaling}.

\end{appendix}


\bibliography{refs}

\end{document}